\documentstyle[prl,aps,multicol,epsfig,psfig,amssymb]{revtex}
\begin{document}
\draft
\title{Dynamics of rough interfaces in Chemical Vapor Deposition: 
experiments and model for silica films}

\author{Fernando Ojeda$^1$, Rodolfo Cuerno$^2$, Roberto Salvarezza$^3$,
and Luis V\'azquez$^1$}
\address{
$^1$ Instituto de Ciencia de Materiales de Madrid (CSIC),
Cantoblanco, E-28049 Madrid, SPAIN \\
$^2$ Departamento de Matem\'aticas and
Grupo Interdisciplinar de Sistemas Complicados, \\
Universidad Carlos III de Madrid, Avenida de la Universidad 30,
E-28911 Legan\'{e}s, SPAIN \\
$^3$ INIFTA, Sucursal 4, Casilla de Correo 16, (1900) La Plata, ARGENTINA}

\date{\today}
\maketitle

\begin{abstract}
We study the surface dynamics of silica films grown by low pressure
chemical vapor deposition. Atomic force microscopy measurements show 
that the surface reaches a scale invariant stationary state compatible
with the Kardar-Parisi-Zhang (KPZ) equation in three dimensions. At 
intermediate times the surface undergoes an unstable transient due to 
shadowing effects.
By varying growth conditions and using spectroscopic techniques, 
we determine the physical origin of KPZ scaling to be a low value of the 
surface sticking probability, related to the surface concentration of reactive 
groups. We propose a stochastic equation that describes
the qualitative behavior of our experimental system.
\end{abstract}


%68.55.Jk: Structure and morphology; thickness
%64.60.Ht: Dynamic critical phenomena
%81.15.Gh: Chemical Vapor Deposition
%05.10.Gg: Stochastic analysis methods (Fokker-Planck, Langevin)

%\tighten

%\narrowtext

\begin{multicols}{2}

The dynamics of growing interfaces has become of increasing interest in 
order to understand the physical processes that determine film quality 
\cite{alb}. In the absence of morphological instabilities, many 
surfaces evolving out of equilibrium display time and space scale invariance, 
$i.e.$ such surfaces are {\em rough}.
A successful framework for the study of rough interface 
dynamics has been the formulation of stochastic differential equations 
for the surface height $h(\bbox{r},t)$, where $\bbox{r}$ denotes a site of
a two-dimensional substrate, and $t$ is time. 
A prominent example is the Kardar-Parisi-Zhang (KPZ) equation \cite{kpz} 
\begin{equation}
\partial_t h = \nu \nabla^2 h +
(\lambda/2) (\nabla h)^2 + \eta(\bbox{r},t),
\label{KPZeq}
\end{equation}
expected to describe the large length scale
dynamics of any rough surface growing in the absence of specific conservation
laws. In Eq.\ (\ref{KPZeq}), $\nu$ and $\lambda$ are constants.
The first term on the rhs is sometimes 
called a surface tension term, as it tends to smooth the interface through
evaporation-condensation processes \cite{alb}. 
The nonlinear term accounts for growth along the local normal direction
(lateral growth), and $\eta(\bbox{r},t)$ is a Gaussian white noise
of constant strength $2D$, accounting for microscopic fluctuations, 
$e.g.$, in a deposition beam.

One of the techniques for thin film production which might be expected to 
lead to surfaces described by (\ref{KPZeq}) is Chemical Vapor Deposition (CVD).
CVD is very widely used for technological applications and industrial devices 
\cite{jk}. It is also very interesting from the fundamental point of view
due to its conceptual similarities with other growth techniques such 
as electrodeposition. Thus, a considerable effort has been made to 
model CVD growth \cite{vdbj,pg,brz} using partial 
differential equations. These models predict the development of 
unstable morphologies within the time ranges studied, contradicting the 
above expectation of KPZ scale invariant behavior. Let us note that, to date,
very few experimental systems are known whose scaling is described
by KPZ in three dimensions \cite{kpzexp}. Moreover, on the fundamental level 
there are no experiments addressing the long time behavior of the surface 
morphologies produced by CVD, and the identification of the determining 
physico-chemical mechanisms.

In this Letter we study the interface dynamics of low pressure SiO$_2$ 
CVD films by Atomic Force Microscopy (AFM), Raman and infrared 
spectroscopies, and numerical simulations. SiO$_2$ films were studied because 
of their amorphous nature, to avoid Schwoebel barrier effects on the 
surface dynamics \cite{sch}, and to prevent formation of facets that can also 
alter the scaling behavior \cite{brz}. Our experimental system displays
a transient unstable behavior related to the gas phase transport character 
of CVD growth \cite{vdbj,pg,brz}, but a stationary state is achieved 
which, rather, is compatible with KPZ scaling. 
The surface dynamics is well described, over a wide spatial 
and temporal range (up to 2 days of deposition time), by a stochastic 
differential equation in which surface diffusion, shadowing effects, 
and lateral growth are allowed for. 

Amorphous SiO$_2$ films were grown at a temperature $T$ = 723 K on Si (100) 
substrates in a hot wall, horizontal, low pressure tubular CVD reactor. 
The precursor gases were silane (diluted at 2\% in nitrogen, 99.999\% purity) 
and oxygen (99.9992\% purity), with an oxygen/silane ratio equal to 20 and 
a total gas flow rate of 50 sccm. The chamber pressure was 1.4 Torr. 
The film thickness increases linearly with deposition time at a constant 
growth rate of 20 nm/min. Films were deposited in the 
range 5 min $\leq t \leq$ 2 days. The surface morphology was characterized by 
AFM (Nanoscope III from Digital Instruments, CA) operating in tapping mode 
at ambient conditions up to a scale of 50 $\mu$m using silicon cantilevers.

AFM imaging of our silica films shows (Fig.\ 1) a deposit formed 
initially by small rounded grains 30-60 nm in size. A similar morphology 
has been reported on amorphous silicon films deposited by thermal evaporation 
\cite{rensse}. As deposition proceeds, structures resembling mountains 
and valleys appear at larger length scales increasing in size, until a 
stationary regime is attained in the 15-30 hours range. 

Power Spectral Density (PSD) (or surface power spectrum) 
plots of the surface morphology at different times are 
displayed in Fig.\ 2. For $t<$ 35 min and small enough wave vector
$k$, the PSD follows 
that of the initial substrate, while for $t>$ 35 min and $k < k_c$ it takes 
a constant $k$-independent value, interface portions separated by 
distances $r >1/k_c$ being uncorrelated. For a rough interface \cite{alb}, 
the value of $k_c$ decreases with time as $k_c \sim t^{-1/z}$,
where $z$ is the {\em dynamic} exponent. 
Indeed, from the log-log plot of $k_c $ vs $t$ (Fig.\ 3), 
a slope value $z = 1.6 \pm 0.1$ is obtained. Therefore, 
the time behavior of our CVD growth process is well described by that of a 
self-affine rough surface. The same conclusion is obtained from the 
spatial behavior of the surface morphology. Specifically, for a 
two-dimensional rough interface the PSD behaves as \cite{alb} PSD$(k) 
\sim 1/k^{2\alpha +2}$, with $\alpha$ the {\em roughness}
exponent. In Fig.\ 2, as time proceeds, up to three spatial regions 
can be observed:

{\em i)} For $t < 50$ min we can distinguish two regions: 
Region I, for $k>k_0$, features $\alpha_I = 0.99 \pm 0.04$; $k_0$ initially 
decreases from 0.04 nm$^{-1}$ for $t=20$ min to a constant value 
of 0.02 nm$^{-1}$ for $t \geq$ 50 min. Region II with $\alpha_{I\!I} =0.75 
\pm 0.03$ is observed for $k_1 < k < k_0$, where $k_1$ changes from 
0.025 nm$^{-1}$ for $t$ = 20 min to 0.0034 nm$^{-1}$ for $t$ = 50 min. 

{\em ii)} For $t > 50$ min a new region III appears for $k_c < k 
< k_1$, with $\alpha_{I\!I\!I} = 0.42 \pm 0.03$; as noted above,
$k_c$ decreases with time. Regions I and II are still observed.

The above behavior of the PSD is compatible with that of the surface 
width, $W^2(t) = \langle \overline{(h(\bbox{r},t) -\overline{h}(t))^2}
\rangle $, which we measure independently (Fig.\ 3). The bar
denotes spatial average. For a rough interface,
initially $W(t) \sim t^{\alpha/z} = t^{\beta}$, 
as long as $t \ll L^z$, with $L$ a
measure of the spatial extent of the system. For $t \gg L^z$, the
interface saturates into a stationary state for which $W \sim L^{\alpha}$.
As a difference with the PSD, we observe only two temporal 
behaviors in $W(t)$, corresponding to regions II and III of the PSD.
For $t <$ 50 min, the width is described by the {\em growth} exponent
$\beta_{I\!I} = 0.42 \pm
0.04$, while for 50 min $ \leq t \leq $ 15-30 hours, $W(t)$ data are
consistent with $\beta_{I\!I\!I} = \alpha_{I\!I\!I}/z = 0.26 \pm 0.03$ as
obtained from the PSD data. For $t >$15-30 hours, the width $W(t)$ 
saturates to a constant value. Note that the contribution of the short 
length scales (region I) to the surface width is masked by the spatial 
region II. However, despite the few data points for $t \leq$ 50 min, 
an attempt to estimate $\beta_I$ from 
the PSD curves \cite{integr} leads to $\beta_I = 0.28 \pm 0.09$. 
Regarding the long distance properties of our system,
the observed $\alpha_{I\!I\!I} = 0.42 \pm 0.03$,
$z = 1.6 \pm 0.1$, and $\beta_{I\!I\!I} = 0.26 \pm 0.03$ are close,
within experimental errors, to the approximate exponents for
the KPZ universality class in three dimensions, namely \cite{alb},
$\alpha_{K\!P\!Z} = 0.39$, $z_{K\!P\!Z} = 1.63$, and $\beta_{K\!P\!Z} =
0.24$. Thus, the asymptotic scaling behavior of the growing CVD surface 
is of the KPZ class, an experimental result very rarely found in 
three dimensions \cite{kpzexp}. 

It is important to understand the origin of the KPZ regime
under our experimental conditions. It is known \cite{ch} that,
for CVD systems with a low surface reaction kinetics
(low sticking probability $s$), lateral growth, which is the
fingerprint of KPZ behavior, is promoted
leading to the so-called conformal growth. The mechanism inducing
conformal growth is the reemission of the precursor ($i.e.$, depositing)
species due to a low sticking probability.
For our experimental conditions $s$ has actually
been reported \cite{wk} to decrease with temperature from $s$ = 0.5 at 573 K
down to $s$ = 0.08 at 723 K. Also, the growth conformality
of submicron trenches by SiO$_2$ films has been reported \cite{wk,ik}
to improve as $T$ increases. Physically, $s$ can be related to the
concentration of reactive sites at the growing film surface, which for our
system are mainly associated with hydrogenated groups,
such as Si-OH and Si-H, and with strained siloxane (Si-O-Si) groups, rather
than with the less reactive relaxed siloxane groups \cite{hydrox}.

To study the role played by the sticking probability on our film morphology,
we have grown SiO$_2$ films at lower $T$ = 611 K (higher $s$) with the
{\em same growth rate} ($\approx$ 20 nm/min). Analysis of our films by
Raman and infrared spectroscopies shows \cite{us}
that the 611 K films indeed present a
significantly higher concentration of reactive groups than the 723 K films,
in agreement with the higher sticking probability expected for the low
temperature growth. Moreover, regarding the surface dynamics of the low
$T$ films, neither the KPZ nor the saturation regimes were found after 2
days of deposition. In Fig.\ 4 we plot the PSD of a film grown
at 611 K after 2 days. The KPZ region does not appear but only
regions I and II do. In Fig.\ 4 (inset) we plot $W(t)$ for the same
set of films. Again, only unstable growth is obtained for long times.
From the point of view of Eq.\ (\ref{KPZeq}), these experiments
suggest that when $T$ is reduced the effective $\lambda$ coefficient of
the KPZ term decreases. These results are analogous
to the crossover found between diffusion-limited aggregation and Eden growth
when tuning the sticking probability \cite{meakin},
and allow us to identify region II in the scaling behavior at the high
tempererature value ($T=723$ K) as an unstable transient.

The above results suggest that the KPZ equation might be a good starting point 
to describe the observed behavior. However, in our system we expect the 
first term in equation (\ref{KPZeq}) to be negligible, since the SiO$_2$ vapor 
pressure is extremely low ($i.e.$ $\nu$ is very small) in our temperature 
range, whereas surface stabilization by surface diffusion seems to be relevant 
\cite{rensse}. In fact, $\alpha_I \approx 0.99$ and $\beta_I \approx 0.28$ are 
consistent within experimental errors with the linear theory of surface 
diffusion \cite{linmbe}. On the other hand, due to the large values of the 
effective exponents $\alpha_{I\!I}$ and $\beta_{I\!I}$ 
and in view of the discussion above, 
we believe region II corresponds to unstable 
growth. This instability can be related to shadowing effects 
which occur in CVD \cite{vdbj,pg,brz} due to the random walk motion of the
depositing particles. 
Nonlinear surface diffusion \cite{fam} can be discarded as the origin
of the unstable behavior in region II since it leads to anomalous scaling 
\cite{alb}, not present in our measurements. Similarly, the unstable 
equation for $h$ derived in \cite{brz} can be ruled out as a description 
of the unstable behavior in region II because it leads to a non constant
growth rate, incompatible with our experimental setup. Thus, we propose 
the following continuum equation to describe the silica CVD growth:
\begin{equation}
\partial_t h = -K \nabla^4 h + \varepsilon \; \theta/\overline{\theta} 
+ (\lambda/2) (\nabla h)^2 + \eta(\bbox{r},t) .
\label{eq}
\end{equation}
In Eq.\ (\ref{eq}), $K$ and $\varepsilon$ are positive constants;
the first term on the rhs represents relaxation by surface 
diffusion and the second term represents geometric shadowing effects, 
$\theta$ being the local exposure angle and 
$\overline{\theta}$ being the spatial average of $\theta$ \cite{shadow}. 
In order to check the validity of (\ref{eq}) we have performed numerical 
simulations in two dimensions
(three dimensional simulations are limited to small system sizes 
\cite{shadow2}). In Fig.\ 5 we plot the time evolution of the PSD 
for $K=D=1$, $\varepsilon=1/2$, and two values of the strength of the
KPZ non-linearity $\lambda=0.2, 3$. 
For these sets of parameters we have employed system sizes up to $L$ = 8192
in order to reliably identify the different scaling regimes.
For $\lambda=3$ (solid line in Fig.\ 5),
and as experimentally observed at high $T$ (Fig.\ 2), the 
surface PSD evolves from a scaling regime dominated by surface diffusion at 
short distances to KPZ scaling at large length scales, 
through an unstable transient
due to the shadowing effects \cite{exp}. Also in agreement with the 
experimental data, after a certain growth time both $k_0$ and $k_1$ are 
frozen, and no anomalous scaling is observed. The time evolution of the surface 
width is shown in the inset of Fig.\ 5 for the same parameters. 
Again, for $\lambda=3$, it resembles quite closely that shown in Fig.\ 3 
\cite{exp} since, as $t$ increases, the behavior changes from surface 
diffusion to unstable growth and finally to KPZ scaling. 
For smaller $\lambda =0.2$ (dashed lines in Fig.\ 5), corresponding
experimentally to low $T$ (Fig.\ 4),
we observe from the PSD and $W(t)$ behaviors that the crossover from the 
unstable region II to KPZ behavior takes place, if at all, at longer 
time and larger length scales. Similar behaviors to those
in Fig.\ 5 are obtained for other parameter sets \cite{us}, 
provided that the relative weight of the different contributions is preserved.

In summary, we have found that low pressure CVD growth of silica films 
is governed by the relative balance between surface diffusion, shadowing and 
lateral (KPZ) growth. Our experimental system is well described by 
a differential continuum stochastic equation in which these three 
mechanisms are allowed for. Moreover, our study allows to link the 
value of the effective KPZ nonlinearity to the physical and
chemical properties of the growing interface. Thus, we conclude that
the observation of asymptotic KPZ scaling is favoured under growth conditions 
($e.g.$, high temperatures) which promote a low sticking probability of 
the depositing species.

This work has been performed within the CONICET-CSIC and 
Programa de Cooperaci\'on con Iberoam\'erica (MEC) research programs, 
and has been partially supported by 
CAM grants 7220-ED/082, 07N-0028, and DGES grants MAT97-0698-C04, PB96-0119. 
F.\ O.\ acknowledges support by CAM.

\narrowtext

\vskip 1.3cm

\begin{figure}
\begin{center}
\epsfig{file=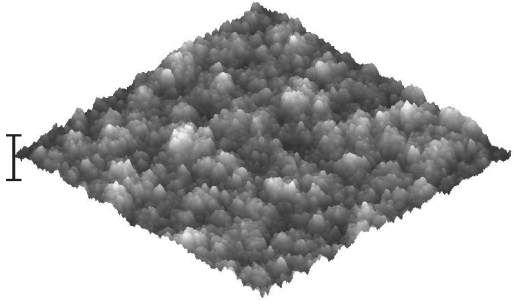, width=2.7in, height=4cm}
\vskip 0.3cm
\caption[]{AFM 3$\times$3 $\mu$m$^2$ image of amorphous SiO$_2$ films
deposited at $T$ = 723 K for $t$ = 75 min. The vertical bar represents 400 nm.}
\end{center}
\end{figure}

\vskip -1.2cm 

\begin{figure}
\begin{center} 
\epsfig{file=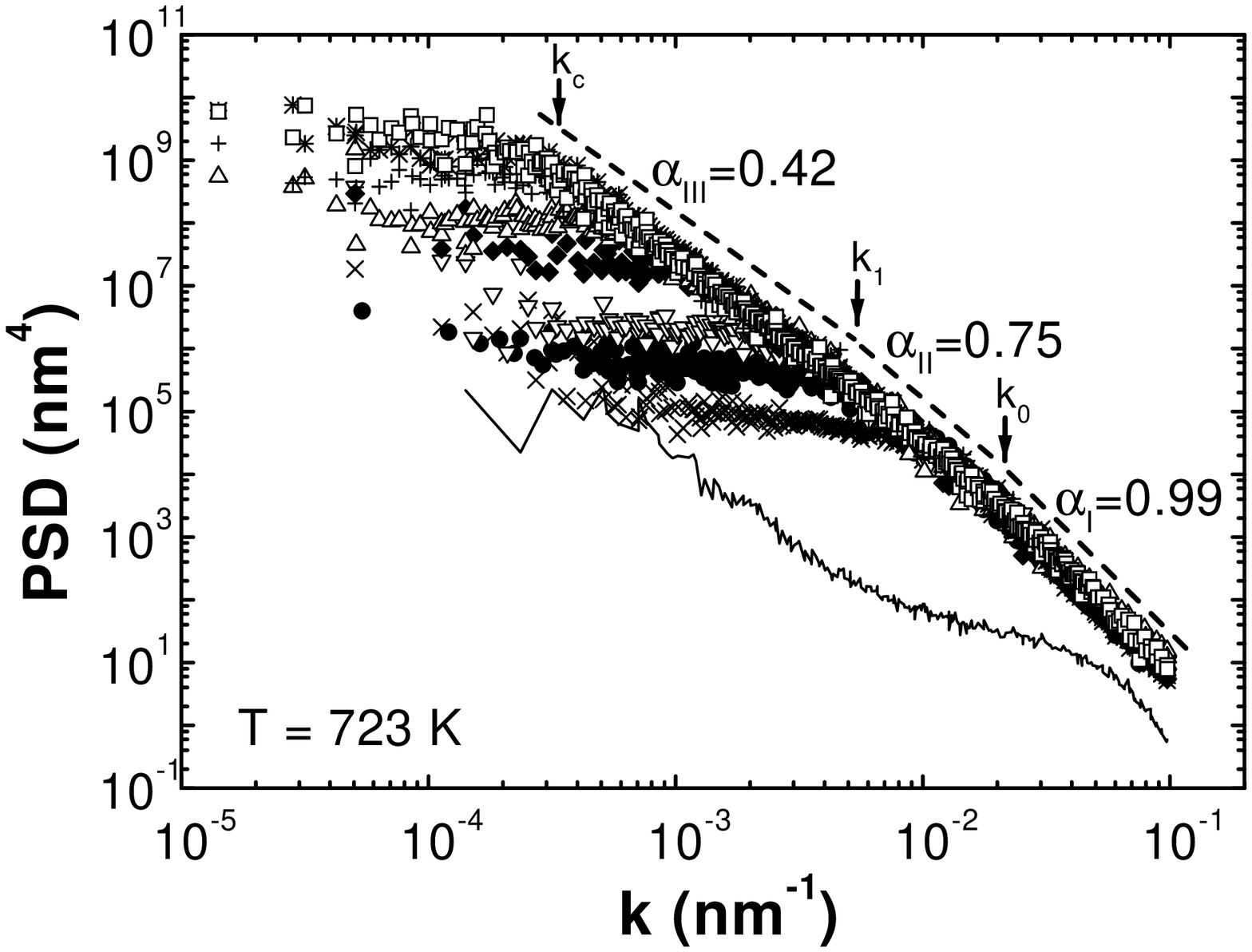, width=3.4in}
\vskip 0.3cm
\caption{PSD curves from AFM images of the SiO$_2$ surface at different times: 
substrate (solid line), 5 min ($\times$), 20 min ($\bullet$), 
50 min ($\triangledown$), 180 min ($\blacklozenge$), 480 min ($\vartriangle$),
900 min ($+$), 1800 min ($*$), and 2880 min ($\Box$). Wave vectors 
$k_0$, $k_1$, and $k_c$ separating the scaling regions are shown 
for $t=2880$ min. The slopes of the dashed lines are
$\alpha_{I\!I\!I} = 0.42$ (for $k_c < k < k_1$), $\alpha_{I\!I}= 0.75$ 
(for $k_1 < k < k_0$), and $\alpha_I = 0.99$ (for $k > k_0$).} 
\end{center}
\end{figure}

\vskip 0.0cm

\begin{figure}
\begin{center}
%\hskip -0.5in
\epsfig{file=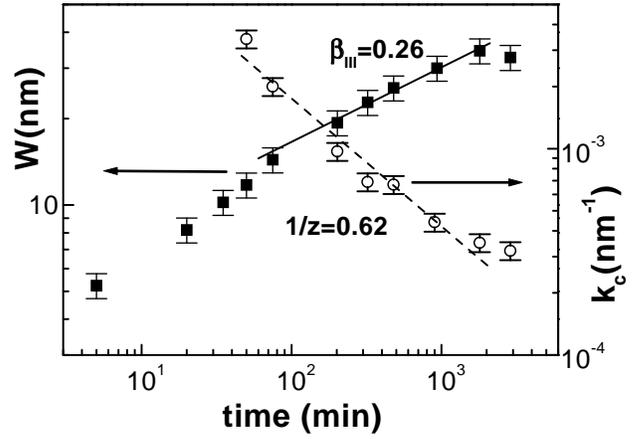, width=3.2in}
\vskip 0.3cm
\caption[]{Plot of $W(t)$ ($\blacksquare$, left axis) 
and $k_c(t)$ ($\circ$, right axis). 
The slopes of the solid and dashed lines are, respectively, 
$\beta_{I\!I\!I} = 0.26$ and $1/z=0.62$.}
\end{center}
\end{figure}

\vskip 0.5cm

\begin{figure}
\begin{center}
%\hskip -1cm
\epsfig{file=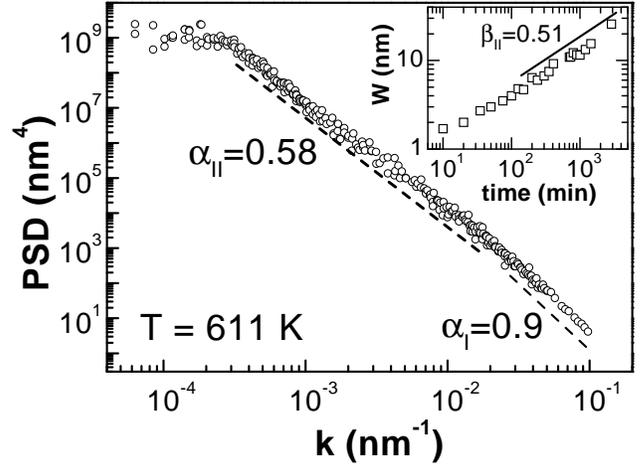, width=3.3in}
\vskip 0.3cm
\caption{PSD of a SiO$_2$ film grown for 2 days at 611 K.
Slopes of the dashed lines are 
$\alpha_{I\!I}=0.58$, $\alpha_I=0.9$. 
Inset: $W(t)$ for the same system.
Slope of the solid line is $\beta_{I\!I}=0.51$.}
\end{center}
\end{figure}

\vskip 0.3cm

\begin{figure}
\begin{center}
\epsfig{file=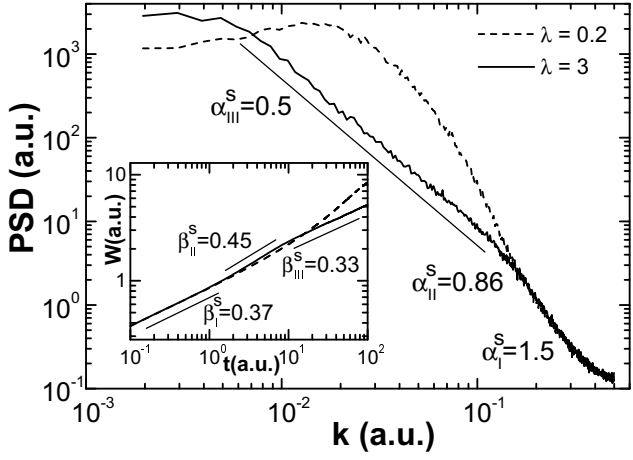, width=3.3in}
\vskip 0.3cm
\caption{PSD from numerical integrations of Eq.\ (\ref{eq})
for $K=D=1, \varepsilon=1/2$, $L= 1024$, $t=100$, averaged over
250 realizations, for $\lambda=0.2$ (dashed line) and $\lambda=3$ (solid line).
Slope of the straight solid line is $\alpha_{I\!I\!I}^s=0.5$.
For $\lambda=3$, the values $\alpha_I^s=1.5$, $\alpha_{I\!I}^s=0.86$ 
give the best fit
to the PSD behavior in each region. Data curvature at large $k$
is due to discretization effects.
Inset: $W(t)$ for the same parameters. Slopes of the straight
solid lines are $\beta_I^s=0.37$, $\beta_{I\!I}^s=0.45$, and
$\beta_{I\!I\!I}^s=0.33$.}
\end{center}
\end{figure}

\end{multicols}


\begin{references}
\bibitem{exp} The values of $\alpha$ and $z$ 
depend on dimensionality, and thus differ in the experiments and in
our numerical simulations. However, the qualitative behaviors are the same: 
for Fig.\ 5 we obtain for the $\lambda=3$ case
$\alpha_I^s = 1.5 \pm 0.1$ 
($\beta_I^s = 0.37 \pm 0.01$), $\alpha_{I\!I}^s= 0.86 \pm 0.04$ 
($\beta_{I\!I}^s = 0.45 \pm 0.02$), and $\alpha_{I\!I\!I}^s = 0.5 \pm 0.01$ 
($\beta_{I\!I\!I}^s = 0.33 \pm 0.02 $), consistent, respectively, 
with linear surface diffusion, unstable growth, and KPZ scaling in two 
dimensions.
\end{references}
\end{document}